\documentclass[12pt]{article}
\usepackage{amssymb}
\begin{document}
\title{\bf Killing and Noether Symmetries of Plane Symmetric Spacetime}
\author{M. Farasat
Shamir\thanks{farasat.shamir@nu.edu.pk}, Adil Jhangeer
\thanks{adil.jhangeer@gmail.com} and Akhlaq Ahmad Bhatti \thanks{akhlaq.ahmad@nu.edu.pk} \\\\ $^{*,~ \ddag}$
National University of Computer \& Emerging Sciences,\\Lahore
Campus, Pakistan\\\\$^{\dag}$ Deanship of Educational Services,
Qassim University,\\Buraidah, Qassim, KSA.}

\date{}
\maketitle
\begin{abstract}
This paper is devoted to investigate the Killing and Noether
symmetries of static plane symmetric spacetime. For this purpose,
five different cases have been discussed. The Killing and Noether
symmetries of Minkwoski spacetime in cartesian coordinates are
calculated as a special case and it is found that Lie algebra of
the Lagrangian is $10$ and $17$ dimensional respectively. The
symmetries of Taub's universe, anti-deSitter universe, self
similar solutions of infinite kind for parallel perfect fluid case
and self similar solutions of infinite kind for parallel dust case
are also explored. In all the cases, the Noether generators are
calculated in the presence of gauge term. All these examples
justify the conjecture that Killing symmetries form a subalgebra
of Noether symmetries \cite{kashif}.
\end{abstract}


\section{Introduction}

Symmetry is an important phenomenon to understand the universe.
Most of the stars are assumed to have sphere like symmetry.
Cylindrical and plane symmetries may be used to investigate the
gravitational waves. Since general relativity (GR) and other
modified theories such as $f(R)$ \cite{fR}, $f(T)$ \cite{fT},
$f(R,T)$ \cite{fRT} theories of gravity, Horva-Lifhitz
\cite{Horva} gravity and Brans-Dicke theory \cite{BD} etc. are
highly non linear, so the solutions of their field equations can
be understood through their symmetries. These symmetries are given
by isometries or Killing vectors (KVs) of the spacetime. An
isometry is a direction along which Lie derivative of the metric
tensor is zero. It would be worthwhile to mention here that many
GR solutions do have some symmetry \cite{Bokhari and Kara}. The
problem of localization of energy momentum in GR can be addressed
using symmetries. There always exists a timelike Killing Vector
(KV) for static spacetimes which may be used to define the energy
of a particle using the equation $E = \alpha. p$, where $\alpha$
is the timelike KV and $p$ represents the momentum of the particle
\cite{kramer}. Similarly the rotational symmetries of static and
spherically symmetric spacetime such as an exterior of a star
provide conservation of angular momentum. Thus symmetries play an
important role in many physical applications of gravitational
fields.

Another systematic approach to find the conserved quantities
\cite{CSC, TTSP} of the variational problems is given by Emmy
Noether. She gave a relationship between the conservation laws and
symmetries \cite{TTSP}, known as Noether symmetries. Conservation
laws are very important in the field of differential equations. They
help in computing the unknown exponent in the similarity solution
which  may not be obtained from the homogeneous boundary conditions
\cite{NMM1}. The conserved quantities are also proved to be helpful
to control numerical errors in the integration process of partial
differential equations (PDEs). The celebrated Noether's theorem
\cite{TTSP} states that for every infinitesimal generator of
Lagrangian symmetry, there exist a conserved quantity. The
invariance of time translation corresponds to the conservation of
energy. Thus it is hoped that the Noether symmetry approach may
become helpful to define the energy content of gravitational waves.
This theorem also suggests that constants of motion for a given
Lagrangian are related to its symmetry transformations
\cite{relate}.

The importance of Noether symmetries can be seen from the fact that
these symmetries may recover some lost conservation laws and
symmetry generators of spacetimes \cite{lost}. Sharif and Saira
\cite{siara1} calculated the energy contents of colliding plane
waves by using approximate symmetries. It has been concluded that
there does not exist any nontrivial first order symmetry for plane
electromagnetic and gravitational waves. The same authors
\cite{siara2} used approximate Lie symmetries to explore the energy
contents of Bardeen model and stringy charged black hole type
solutions. In a recent paper \cite{shamir}, we have investigated
$f(R)$ theory of gravity using Noether symmetries in the presence of
gauge term. For this purpose, we discussed Noether symmetry
generators for spherically symmetric spacetimes and Friedmann
Robertson Walker universe along with the importance and stability
criteria of some well known $f(R)$ gravity models. Lie symmetry
methods for differential equations were used to explore the
symmetries of a perturbed Lagrangian for a plane symmetric
gravitational wave like spacetime \cite{iqadir}. In recent years,
many authors have studied Noether and Killing symmetries in
different contexts.

Hussain et al. \cite{Hussain} investigated second order approximate
symmetries of the geodesic equations for the Reissner-Nordström
metric and it was concluded that energy must be re-scaled. The same
authors \cite{Hussain1} gave a  proposal to determine the energy
content of gravitational waves using approximate symmetries of
differential equations. They also investigated exact and approximate
Noether symmetries of the geodesic equations for the charged Kerr
spacetime in \cite{Hussain2}. The re-scaling of the energy contents
was done and it was conjectured that for any spacetime, the
conformal KVs form a subalgebra of the symmetries of the Lagrangian
that minimizes arc length. Later on, Hussain \cite{Hussain3} gave a
counter example to prove that the conjecture was not true in general
for any spacetime with non-zero curvature. Bokhari et al.
\cite{kashif} studied the symmetry generators of Killing vectors and
Noether symmetries. It was concluded that the Noether symmetries
contained the set of Killing symmetries. Thus they gave a conjecture
that the Noether symmetries form  a bigger set of Lie algebra and
the Killing symmetries are a subset of that algebra. A similar
comparison of Killing and Noether symmetries was given for
conformally flat Friedmann metric \cite{kara}. Hall \cite{hall} gave
a general relation connecting the dimensions of the Killing algebra,
the orbits and the isotropies of the spacetime.

A plane symmetric spacetime may be considered as a Lorentzian
manifold which possess a physical energy momentum tensor. It admits
the the minimal isometry group $SO(2)$ x $\mathbb{R}^2$ such that
the group orbits form spacelike surfaces with constant curvature.
Feroze et al. \cite{Feroze} discussed a complete classification of
plane symmetric Lorentzian manifolds according to their additional
isometries by solving the Killing equations. Classification of
static plane symmetric spacetimes according to their matter
collineations is given by Sharif \cite{sharifplane}. In this paper,
we are focussed to investigate the Killing and Noether symmetries of
plane symmetric spacetime. For this purpose we take five different
cases namely Minkwoski spacetime (as a special case), Taub's
universe, anti-deSitter universe, self similar solutions of infinite
kind for parallel perfect fluid and self similar solutions of
infinite kind for parallel dust case. The plan of the paper is as
follows: In second section, we give some basics of KVs and Noether
symmetries. Sections $3$ and $4$ are used to calculate the Killing
and Noether symmetries of the spacetimes mentioned above. In the
last section we summarize and compare the results.

\section{Killing and Noether Symmetries}

A transformation that leaves the object (in our case, the metric)
invariant is called symmetry. However, the precise definition of the
symmetry is as follows: A symmetry of an object is a diffeomorphic
mapping of the object to itself that preserves the structure of the
object and leaves it invariant. Now let the mapping $T$ of an object
$P$ be
\begin{eqnarray*}
T(\lambda): P\rightarrow P,
\end{eqnarray*}
where $\lambda$ parameterizes the symmetry. For example, the
Minkwoski spacetime is time invariant and its translational
invariance is given by
\begin{eqnarray*}
T(\lambda): (t,x^i)\rightarrow (t+\lambda,x^i).
\end{eqnarray*}
Mainly there exist two types of transformations, discrete and
continuous. Reflection is an example of discrete transformation
while translations and rotations are continuous transformations. In
GR, continuous transformations are most important as these can be
obtained in a systematical and mathematical way by finding the
Killing vectors of a spacetime. The transformation which leaves the
metric tensor $g_{ij}$ invariant, is called an isometry. Thus an
isometry is a direction along which the metric tensor is Lie
transported, i.e. if the vector field $K$ is an isometry then
\begin{eqnarray*}
\pounds_K g_{ij}=0,
\end{eqnarray*}
where for a four dimensional spacetime, $i,j=0,1,2,3$. In index
notation, we have
\begin{eqnarray*}
K_{i;j}+K_{j;i}=0,
\end{eqnarray*}
where semi colin denotes covariant derivative and this equation is
known as Killing's equation. Any vector satisfying this equation is
called Killing vector. Thus $K$ is a Killing vector field if and
only if
\begin{eqnarray*}
K_{(i;j)}=0.
\end{eqnarray*}
It is mentioned here that the symmetries of a manifold may be
characterized by its KVs and a finite dimensional Lie group is
formed \cite{Hawking}.

Noether symmetries are also known as symmetries of a Lagrangian.
The method for calculating the Noether gauge symmetries by using
the Lagrangian is given below.
\\\\Let
\begin{equation}
ds^{2}= g_{ij} dx^{i}dx^{j} \label{T1}
\end{equation}
be a line element then the vector field $X$ for (\ref{T1}) will be
\begin{equation}
X=\xi(s,x^{i}){\frac{\partial}{\partial{s}}}+\eta^{j}(s,x^{i}){\frac{\partial}{\partial{x^{j}}}}.
\label{T2}
\end{equation}
The Lagrangian $L$ for (\ref{T1}) can be computed by using
\begin{equation}
L= \frac{1}{2}g_{ij}{\dot{x}^{i}}{\dot{x}^{j}}, \label{T3}
\end{equation}
where dot denotes derivative with respect to $s$. The Noether
gauge symmetry is given by the equation
\begin{equation}\label{noether}
X^{[1]}(L)+ L D_{s}(\xi)=D_{s}(G), \label{T4}
\end{equation}
where $G$ is a gauge function and $X^{[1]}$ is the first
prolongation given by
\begin{equation}
X^{[1]}=X + (\eta^{j}_{~,s}+\eta^{j}_{~,i}{\dot{x}}^i-
\xi_{,s}{\dot{x}}^j-\xi_{,i}{\dot{x}}^i{\dot{x}}^j)\frac{\partial}{\partial
\dot{x}^j}
\end{equation}
and $D_{s}$ is defined as
\begin{eqnarray}
D_{s}={\frac{\partial}{\partial{s}}}+{\dot{x}^{i}}{\frac{\partial}{\partial{x}^{i}}}.
\nonumber
\end{eqnarray}

\section{Killing Symmetries of Plane Symmetric spacetime}

The general static plane symmetric spacetime is given by
\cite{kramer}
\begin{equation}
ds^{2}=A(x)dt^{2}-C(x)dx^{2}-B(x)(dy^{2}+dz^{2}).
\end{equation}
For simplicity we take the coefficient of $dx^{2}$ equal to unity
so that the above spacetime becomes
\begin{equation}\label{33}
ds^{2}=A(x)dt^{2}-dx^{2}-B(x)(dy^{2}+dz^{2}).
\end{equation}
For (\ref{33}), the Killing equations turn out to be
\begin{eqnarray}\label{k1}
{k^1}_{,x}=0,\\
{k^2}_{,z}+{k^3}_{,y}=0,\\
A_{,x}k^1+2A{k^0}_{,t}=0,\\
A{k^0}_{,x}-{k^1}_{,t}=0,\\
A{k^0}_{,y}-B{k^2}_{,t}=0,\\
A{k^0}_{,z}-B{k^3}_{,t}=0,\\
{k^1}_{,y}+B{k^2}_{,x}=0,\\
{k^1}_{,z}+B{k^3}_{,x}=0,\\
B_{,x}{k^1}+2B{k^2}_{,y}=0,\\\label{k10}
B_{,x}{k^1}+2B{k^3}_{,z}=0.
\end{eqnarray}
We suppose the vector field $X$ of the form
\begin{equation}
X=k^0(t,x,y,z)\frac{\partial}{\partial
t}+k^1(t,x,y,z)\frac{\partial}{\partial x}+
k^2(t,x,y,z)\frac{\partial}{\partial
y}+k^3(t,x,y,z)\frac{\partial}{\partial z}.
\end{equation}
Now we solve these equations simultaneously for different values of
metric coefficients $A(x)$ and $B(x)$. We take Minkwoski spacetime
as a special case when $A(x)=B(x)$. The second case is  the Taub's
universe \cite{30} in which $A(x)=x^{-2/3}$ and $B(x)=x^{4/3}$. The
third case gives the anti-deSitter universe \cite{Feroze} for
$A(x)=e^{2x}=B(x)$. The last two cases are self similar solutions of
infinite Kind for parallel perfect fluid case and dust case
\cite{31} when $A(x)=1,~ B(x)=e^{2x}$ and $A(x)=x^2,~ B(x)=1$
respectively.

\subsection{Minkwoski Spacetime}

Here we shall investigate the Killing symmetries by solving Eqs.
(\ref{k1})-(\ref{k10}) for $A(x)=B(x)=1$. After some tedious
calculations, we obtain the solution
\begin{eqnarray*}
k^0&=&c_1x+c_7y+c_8z+c_{10},\\
k^1&=&c_1t+c_4y+c_2z+c_3,\\
k^2&=&c_7t-c_4x+c_5z+c_6,\\
k^3&=&c_8t-c_2x-c_5y+c_9.
\end{eqnarray*}
Thus the Killing symmetries turn out to be
\begin{eqnarray}\label{k13}
X_0=\frac{\partial}{\partial t},~~~~~~X_1=\frac{\partial}{\partial
x},~~~~~~X_2=\frac{\partial}{\partial
y},~~~~~~X_3=\frac{\partial}{\partial z},\\\label{k14}
X_4=y\frac{\partial}{\partial x}-x\frac{\partial}{\partial
y},~~X_5=z\frac{\partial}{\partial y}-y\frac{\partial}{\partial
z},~~X_6=z\frac{\partial}{\partial x}-x\frac{\partial}{\partial
z},\\X_7=x\frac{\partial}{\partial t}+t\frac{\partial}{\partial
x},~~~~X_8=y\frac{\partial}{\partial t}+t\frac{\partial}{\partial
y},~~~~X_9=z\frac{\partial}{\partial t}+t\frac{\partial}{\partial
z},
\end{eqnarray}\\
where these symmetries represent translations in $t,~x,~y,~z$,
rotations in $xy,~yz,~zx$ axis and Lorentz rotations in $x,~y,~z$
respectively. These symmetry generators provides conservation laws
for energy, spin angular momentum and linear momentum \cite{32}.
Minkowski spacetime forms maximal algebra having 10 Killing
symmetries which is known as Poincar´e algebra $so(1,3)$ $\oplus$
$\mathbb{R}^4$, where $\oplus$ represents a semi direct sum
\cite{kramer}.

\subsection{Taub's Universe}

Now we explore the Taub's universe \cite{abc}. In this case the
simultaneous solutions of Eqs.(\ref{k1})-(\ref{k10}) for
$A(x)=x^{-2/3}, ~~B(x)=x^{4/3}$ becomes
\begin{eqnarray*}
k^0&=&c_1,~~~~~~~~~~ k^1=0,\\
k^2&=&c_2z+c_3,~~ k^3=c_4-c_2y.
\end{eqnarray*}
Here we obtain four Killing symmetries $X_0,~X_2,~X_3$ and $X_5$.
It is mentioned here that these four independent KVs provide the
minimal algebra associated with the plane symmetric spacetimes
\cite{kramer}.

\subsection{Anti-deSitter Universe}

For anti-deSitter universe, the solution of the killing equations
turns out to be
\begin{eqnarray*}
k^0&=&\frac{1}{2}\bigg(-e^{-2x}-t^2-y^2-z^2\bigg)c_1-c_2tz-c_3t-c_4ty+c_5y+c_8z+c_{10},\\
k^2&=&\frac{1}{2}\bigg(e^{-2x}-t^2-y^2+z^2\bigg)c_4-c_1ty-c_3y-c_2zy+c_5t+c_7z+c_{6},\\
k^3&=&\frac{1}{2}\bigg(e^{-2x}-t^2+y^2-z^2\bigg)c_2-c_1tz-c_3z-c_4zy+c_8t-c_7y+c_9,\\
k^1&=&c_1t+c_4y+c_2z+c_3.
\end{eqnarray*}
Here we get $10$ Killing symmetries as follows:
\begin{eqnarray}
&&X_0,~~~~~~X_2,~~~~~~~X_3,~~~~~~~X_5,~~~~~~X_8,~~~~~~X_9,\\
\label{kt16}&&X_{10}=\frac{\partial}{\partial x}-
t\frac{\partial}{\partial
t}-y\frac{\partial}{\partial y}-z\frac{\partial}{\partial z},\\
&&X_{11}=\frac{1}{2}\bigg(-e^{-2x}-t^2-y^2-z^2\bigg)\frac{\partial}{\partial
t}+ t\bigg(\frac{\partial}{\partial x}- y\frac{\partial}{\partial
y}-z\frac{\partial}{\partial
z}\bigg),\\&&X_{12}=\frac{1}{2}\bigg(e^{-2x}-t^2-y^2+z^2\bigg)\frac{\partial}{\partial
y}+ y\bigg(\frac{\partial}{\partial x}- t\frac{\partial}{\partial
t}-z\frac{\partial}{\partial
z}\bigg),\\&&X_{13}=\frac{1}{2}\bigg(e^{-2x}-t^2+y^2-z^2\bigg)\frac{\partial}{\partial
z}+ z\bigg(\frac{\partial}{\partial x}- y\frac{\partial}{\partial
y}-t\frac{\partial}{\partial t}\bigg).
\end{eqnarray}

\subsection{Self Similar Solution of Infinite Kind for Parallel Perfect Fluid
Case}

Here the solution becomes
\begin{eqnarray*}
k^0&=&c_1,\\
k^1&=&c_3y+c_7z+c_2,\\k^2&=&\frac{1}{2}\bigg(e^{-2x}-y^2+z^2\bigg)c_3-c_2y-c_4z-c_7zy+c_{6},\\
k^3&=&\frac{1}{2}\bigg(e^{-2x}+y^2-z^2\bigg)c_7-c_2z-c_3zy+c_4y+c_{5}.
\end{eqnarray*}
In this case, we obtain $7$ Killing symmetries which are given by
\begin{eqnarray}
\label{kt13} &&X_0,~~~~~~X_2,~~~~~~X_3,~~~~~~~~X_5,\\
\label{kt16} &&X_{14}=\frac{\partial}{\partial
x}-y\frac{\partial}{\partial y}-z\frac{\partial}{\partial z},\\
&&X_{15}=\frac{1}{2}\bigg(e^{-2x}-y^2+z^2\bigg)\frac{\partial}{\partial
y}+ y\bigg(\frac{\partial}{\partial x}-z\frac{\partial}{\partial
z}\bigg),\\&&X_{16}=\frac{1}{2}\bigg(e^{-2x}+y^2-z^2\bigg)\frac{\partial}{\partial
z}+ z\bigg(\frac{\partial}{\partial x}- y\frac{\partial}{\partial
y}\bigg).
\end{eqnarray}

\subsection{Self Similar Solution of Infinite Kind for Parallel Dust Case}

This case yields the solution of the Killing equations
\begin{eqnarray*}
k^0&=&\frac{1}{x}\bigg[e^{-t}\bigg(c_1y+c_3z+c_5\bigg)-e^{t}\bigg(c_2y+c_4z+c_6\bigg)+c_{10}x\bigg],\\
k^1&=&e^{-t}\bigg(c_1y+c_3z+c_5\bigg)+e^{t}\bigg(c_2y+c_4z+c_6\bigg),\\
k^2&=&-e^{-t}c_1x-e^{t}c_2x+c_7z+c_8,\\
k^3&=&-e^{-t}c_3x-e^{t}c_4x-c_7y+c_9,
\end{eqnarray*}
The corresponding Killing symmetries are:
\begin{eqnarray}
&&X_0,~~~~~~X_2,~~~~~~X_3,~~~~~~~~X_5,\\
&&X_{17}=e^{-t}\bigg[\frac{1}{x}\frac{\partial}{\partial t}+
\frac{\partial}{\partial
x}\bigg],\\&&X_{18}=-e^{t}\bigg[\frac{1}{x}\frac{\partial}{\partial
t}-\frac{\partial}{\partial x}\bigg],\\
&&X_{19}=e^{-t}\bigg[\frac{y}{x}\frac{\partial}{\partial t}+
y\frac{\partial}{\partial x}- x\frac{\partial}{\partial
y}\bigg],\\&&X_{20}=e^{-t}\bigg[\frac{z}{x}\frac{\partial}{\partial
t}+ z\frac{\partial}{\partial x}- x\frac{\partial}{\partial
z}\bigg],\\&&X_{21}=-e^{t}\bigg[\frac{y}{x}\frac{\partial}{\partial
t}-y\frac{\partial}{\partial x}+ x\frac{\partial}{\partial
y}\bigg],\\&&X_{22}=-e^{t}\bigg[\frac{z}{x}\frac{\partial}{\partial
t}-z\frac{\partial}{\partial x}+ x\frac{\partial}{\partial z}\bigg].
\end{eqnarray}
Now we investigate the corresponding Noether symmetries for all
the cases mention above.

\section{Noether Symmetries of Plane Symmetric Spacetime}

The Lagrangian for the Plane symmetric spacetime (\ref{33}) is
\begin{eqnarray}\label{N1}
L=A(x){\dot{t}}^{2}-{\dot{x}}^{2}-B(x)({\dot{y}}^{2}+{\dot{z}}^{2}).\label{c6}
\end{eqnarray}
Using Eq.(\ref{noether}) and after separation of monomials, the
determining equations turn out to be
\begin{eqnarray}\label{N2}
\xi_{,t}=0,~~~\xi_{,x}=0,~~~\xi_{,y}=0,~~~\xi_{,z}=0,\\
\xi_{,s}-2{\eta^1}_{,x}=0,~~~{\eta^2}_{,z}+{\eta^3}_{,y}=0,~~~2{\eta^1}_{,s}+G_{,x}=0,\\
A[2{\eta^0}_{,t}-\xi_{,s}]+A_{,x}\eta^1=0,\\
B[2{\eta^2}_{,y}-\xi_{,s}]+B_{,x}\eta^1=0,\\
B[2{\eta^3}_{,z}-\xi_{,s}]+B_{,x}\eta^1=0,\\
A{\eta^0}_{,x}-{\eta^1}_{,t}=0,~~~A{\eta^0}_{,y}-B{\eta^2}_{,t}=0,~~~A{\eta^0}_{,z}-B{\eta^3}_{,t}=0,\\
{\eta^1}_{,y}+B{\eta^2}_{,x}=0,~~~{\eta^1}_{,z}+B{\eta^3}_{,x}=0,\\
2B{\eta^2}_{,s}+G_{,y}=0,~~~2B{\eta^3}_{,s}+G_{,z}=0,\\\label{N3}
2A{\eta^0}_{,s}-G_{,t}=0,~~~~~~G_{,s}=0.
\end{eqnarray}
This is a system of $19$ differential equations with six unknowns
namely $\xi,~\eta^0,~\eta^1,~\eta^2,~\eta^3$ and $G$. Now we
investigate the Noether symmetries in the presence of gauge term by
solving the above differential equations for the cases discussed in
the previous section.

\subsection{Minkwoski Spacetime}

For Minkwoski spacetime the Lagrangian becomes
\begin{eqnarray}
L={\dot{t}}^{2}-{\dot{x}}^{2}-{\dot{y}}^{2}-{\dot{z}}^{2}.
\end{eqnarray}
Solving Eqs.(\ref{N2})-(\ref{N3}) simultaneously by taking
$A(x)=B(x)=1$, we obtain
\begin{eqnarray*}
\xi&=&\frac{1}{2}c_1s^2+c_2s+c_3,\\
\eta^{0}&=&\frac{1}{2}[(c_1t+c_7)s+c_2t]+c_9x+c_{15}y+c_{16}z+c_{18},\\
\eta^{1}&=&\frac{1}{2}[(c_1x-c_4)s+c_2x]+c_9t+c_{12}y+c_{10}z+c_{11},\\
\eta^{2}&=&\frac{1}{2}[(c_1y-c_6)s+c_2y]+c_{15}t-c_{12}x+c_{13}z+c_{14},\\
\eta^{3}&=&\frac{1}{2}[(c_1z-c_8)s+c_2z]+c_{16}t-c_{10}x-c_{13}y+c_{17},\\
G&=&\frac{1}{2}(t^2-x^2-y^2-z^2)c_1+c_7t+c_4x+c_{6}y+c_8z+c_{5}.
\end{eqnarray*}
In this case we get $17$ Noether symmetries. It is an interesting
fact that $10$ symmetries are exactly the same as the Killing
symmetries while the remaining $7$ are
\begin{eqnarray}\label{N13}
&&X_{23}=\frac{\partial}{\partial
s},~~~X_{24}=s\frac{\partial}{\partial
s}+\frac{1}{2}\bigg(t\frac{\partial}{\partial
t}+x\frac{\partial}{\partial x}+y\frac{\partial}{\partial
y}+z\frac{\partial}{\partial z}\bigg),\\\label{N14}
&&X_{25}=\frac{1}{2}s\bigg(s\frac{\partial}{\partial
s}+t\frac{\partial}{\partial t}+x\frac{\partial}{\partial
x}+y\frac{\partial}{\partial y}+z\frac{\partial}{\partial
z}\bigg),~~~X_{26}=\frac{1}{2}s\frac{\partial}{\partial
t},\\\label{N15} &&X_{27}=-\frac{1}{2}s\frac{\partial}{\partial
x},~~X_{28}=-\frac{1}{2}s\frac{\partial}{\partial
y},~~X_{29}=-\frac{1}{2}s\frac{\partial}{\partial z}.
\end{eqnarray}
Thus we get additional symmetries of spacetime using Noether
approach. Symmetry generator $X_{23}$ indicates translation in $s$
and it is always present in the case of the Lagrangian of minimizing
the arc length \cite{Qadir} while $X_{24}$ represents a scaling
symmetry in $s,~ t,~x,~y$ and $z$ direction. This symmetry is an
important one as it may be used to eliminate the $s$ dependence in
the generators given in Eq.(\ref{N14}) and Eq.(\ref{N15}). It is
mentioned here that Hussain et al. \cite{Hussain} investigated
Noether symmetries of Minkwoski spacetime but with spherical polar
coordinates and obtained a $17$ dimensional Lie algebra.

\subsection{Taub's Universe}

The Lagrangian  for the Taub's spacetime takes the form
\begin{eqnarray}
L=x^{-2/3}{\dot{t}}^{2}-{\dot{x}}^{2}-x^{4/3}[{\dot{y}}^{2}+{\dot{z}}^{2}].
\end{eqnarray}
The simultaneous solution of differential equations
(\ref{N2})-(\ref{N3}) takes the form
\begin{eqnarray*}
\xi&=&c_1s+c_2,~~~\eta^0=\frac{2}{3}c_1t+c_6,~~
\eta^1=\frac{1}{2}c_1x,\\
\eta^2&=&\frac{1}{6}c_1y+c_3z+c_4,~~
\eta^3=\frac{1}{6}c_1z-c_3y+c_5
\end{eqnarray*}
and the gauge term turns out to be constant here which can be taken
as zero without the loss of any generality. This solution forms a
six dimensional algebra. Here $4$ symmetries are exactly the same as
Killing symmetries while the $2$ additional symmetries are
\begin{eqnarray}\label{N16}
X_{23}=\frac{\partial}{\partial
s},~~~X_{30}=s\frac{\partial}{\partial
s}+\frac{2}{3}t\frac{\partial}{\partial
t}+\frac{1}{2}x\frac{\partial}{\partial
x}+\frac{1}{6}y\frac{\partial}{\partial
y}+\frac{1}{6}z\frac{\partial}{\partial z}.
\end{eqnarray}
As before, these two additional symmetries are translation in $s$
and scaling in $s,~ t,~x,~y$ and $z$.

\subsection{Anti-deSitter Universe}

For anti-deSitter universe, the Lagrangian  is given by
\begin{eqnarray}
L=e^{2x}[{\dot{t}}^{2}-{\dot{y}}^{2}+{\dot{z}}^{2}]-{\dot{x}}^{2}.
\end{eqnarray}
In this case, the solution of determining equations becomes
\begin{eqnarray*}
\xi&=&c_1,~~~~~~~~~~~\eta^1=c_2t+c_3z+c_5y+c_4,\\
\eta^0&=&\frac{1}{2}(-e^{-2x}-t^2-y^2-z^2)c_2-c_3zt-c_5yt-c_4t+c_6y+c_9z+c_{11},\\
\eta^2&=&\frac{1}{2}(e^{-2x}-t^2-y^2+z^2)c_5-c_2ty-c_3zy-c_4y+c_6t+c_8z+c_7,\\
\eta^3&=&\frac{1}{2}(e^{-2x}-t^2+y^2-z^2)c_3-c_2tz-c_5yz-c_4z+c_9t-c_8y+c_{10}.
\end{eqnarray*}
The guage term is also zero here. This solution forms $11$
dimensional Lie algebra. In this case, the Killing algebra of
anti-deSitter universe is also a subalgebra of Noether symmetries.
We obtain only one additional symmetry here which is a translation
in $s$, i.e. ${\partial}/{\partial s}$.

\subsection{Self Similar Solution of Infinite Kind for Parallel Perfect Fluid Case}

Here the Lagrangian takes the form
\begin{eqnarray}
L={\dot{t}}^{2}-{\dot{x}}^{2}-e^{2x}[{\dot{y}}^{2}+{\dot{z}}^{2}].
\end{eqnarray}
Thus the solution of the determining equations becomes
\begin{eqnarray*}
\xi&=&c_1,~~~~\eta^0=\frac{1}{2}c_2s+c_{10},\\
\eta^1&=&c_6y+c_4z+c_5,~~~G=c_2t+c_3,\\
\eta^2&=&\frac{1}{2}(e^{-2x}-y^2+z^2)c_6-c_5y-c_7z-c_4zy+c_9,\\
\eta^3&=&\frac{1}{2}(e^{-2x}+y^2-z^2)c_4-c_5z-c_6zy+c_7y+c_8.
\end{eqnarray*}
In this case the gauge function $G$ is a linear function of time.
This solution forms a $9$ dimensional Lie algebra. Out of nine,
seven symmetries are same as the killing symmetries while the
additional symmetries are
\begin{eqnarray*}
X_{23}=\frac{\partial}{\partial s}~~~~~~~ \mathrm{and}~~~~~~~
X_{31}=\frac{s}{2}\frac{\partial}{\partial t}.
\end{eqnarray*}

\subsection{Self Similar Solution of Infinite Kind for Parallel Dust Case}

The Lagrangian for self similar solution of infinite kind for
parallel dust case is
\begin{eqnarray}
L=x^{2}{\dot{t}}^{2}-{\dot{x}}^{2}-{\dot{y}}^{2}-{\dot{z}}^{2}.
\end{eqnarray}
This case yields the solution of the determining equations
\begin{eqnarray*}
\xi&=&\frac{1}{2}c_1s^2+c_2s+c_3,\\
\eta^0&=&\frac{1}{x}[e^{-t}(-c_8\frac{s}{2}+c_{10}y+
c_{11}z+c_{13})-e^{t}(c_9y+c_{12}z-c_7\frac{s}{2}+c_{14})+c_{18}x],\\
\eta^1&=&e^{-t}(-c_8\frac{s}{2}+c_{10}y+c_{11}z+c_{13})+
e^{t}(-c_7\frac{s}{2}+c_9y+c_{12}z+c_{14})+\frac{1}{2}x(c_1s+c_2),\\
\eta^2&=&\frac{1}{2}s(c_1y-c_4)-c_{10}xe^{-t}-c_9xe^{t}+\frac{1}{2}yc_2+c_{15}z+c_{16},\\
\eta^3&=&\frac{1}{2}s(c_1z-c_6)-c_{11}xe^{-t}-c_{12}xe^{t}+\frac{1}{2}zc_2-c_{15}y+c_{17}
\end{eqnarray*}
and the gauge term turns out to be
\begin{eqnarray*}
G=-\frac{1}{2}(x^2+y^2+z^2)c_1+c_4y+c_6z+c_5+c_7xe^{t}+c_8xe^{-t}.
\end{eqnarray*}
This solution also yields a $17$ dimensional Lie algebra in which
$10$ symmetries are same as Killing symmetries while the remaining
$7$ are
\begin{eqnarray}\label{N103}
&&X_{23}=\frac{\partial}{\partial
s},~~~X_{32}=s\frac{\partial}{\partial
s}+\frac{1}{2}(x\frac{\partial}{\partial
x}+y\frac{\partial}{\partial y}+z\frac{\partial}{\partial
z}),\\\label{N104} &&X_{33}=\frac{1}{2}s(s\frac{\partial}{\partial
s}+x\frac{\partial}{\partial x}+y\frac{\partial}{\partial
y}+z\frac{\partial}{\partial z}),\\\label{N105}
&&X_{28}=-\frac{s}{2}\frac{\partial}{\partial
y},~~~X_{29}=-\frac{s}{2}\frac{\partial}{\partial z},\\&&X_{34}=
-\frac{se^{-t}}{2}(\frac{1}{x}\frac{\partial}{\partial t}+
\frac{\partial}{\partial x}),~~~X_{35}=
\frac{se^{t}}{2}(\frac{1}{x}\frac{\partial}{\partial t}-
\frac{\partial}{\partial x}).
\end{eqnarray}

\section{Summary and Conclusion}

The main purpose of this paper is to investigate Killing and Noether
symmetries of static plane symmetric spacetime. For this purpose we
have considered five different cases namely Minkwoski spacetime,
Taub's universe, anti-deSitter universe and self similar solutions
of infinite kind for parallel perfect fluid and dust cases. The
Killing and Noether symmetries for each case are given in the
following table.

\begin{center} \textbf{Table $1$. Comparison of Killing and Noether Symmetries}\\
\vspace{0.2in}
\begin{tabular}{|c|c|c|c|}
\hline Metric& Killing & Noether \\ \hline Minkwoski Spacetime&
$10$& $17$  \\ \hline
Taub's Universe& $4$& $6$  \\ \hline Anti-deSitter Universe& $10$& $11$  \\
\hline Self Similar Solution (Perfect Fluid Case)& $7$& $9$  \\
\hline Self Similar Solution (Dust Case)& $10$& $17$  \\
\hline
\end{tabular}
\end{center}
\vspace{0.2in} It is mentioned here that in all the cases, Killing
symmetries are contained in Noether symmetries. Thus we obtain
additional symmetries of the above mentioned spacetimes by finding
the Noether generators. Moreover, the scalar curvature is constant
(zero and non-zero) in all cases. Thus we conclude that the
Killing symmetries of static plane symmetric spacetime always form
a sub algebra of the Noether symmetries. These examples not only
provides Killing and Noether symmetries but also validate the
conjecture given by \cite{kashif}.
\\\\\textbf{Acknowledgement}\\\\ MFS is thankful to National University
of Computer and Emerging Sciences (NUCES) Lahore Campus, for
funding the PhD programme.

{}
\end{document}